\documentclass[final,,twoside]{IEEEtranTCOM}
\normalsize
\ifCLASSINFOpdf
\else
\fi
\usepackage{fancyhdr}
\usepackage{amsmath,amssymb}
\usepackage{amsfonts}
\usepackage{amsthm}
\usepackage[dvips]{graphicx}
\usepackage{dsfont}
\usepackage{balance}
\usepackage{algpseudocode}
\usepackage{algorithm}

\usepackage{comment}

\usepackage[hidelinks]{hyperref}    

\usepackage{color}
\usepackage{pgfplots}           
\pgfplotsset{compat=1.16}
\usepackage{subfigure}
\usepackage{caption}

\usepackage{enumerate}
\usepackage{graphics}
\usepackage{cite}
\usepackage{mathrsfs}
\usepackage{caption}
\usepackage{stfloats}
\usepackage{xfrac}

\usepackage{tikz}
\usepackage{mathdots}
\usepackage{yhmath}

\usepackage{array}
\usepackage{multirow}
\usepackage{textcomp}
\usepackage{gensymb}
\usepackage{tabularx}
\usepackage{booktabs}
\usepackage{graphicx}

\usepackage{balance}

\let\labelindent\relax

\usepackage{enumitem}

\newtheorem{theorem}{Theorem}

\newtheorem{lemma}{Lemma}

\newtheorem{proposition}{Proposition}
\newtheorem{remark}{Remark}

\begin{document}

\title{Optical RIS-enabled Multiple Access Communications}


\author{Georgios D. Chondrogiannis,~\IEEEmembership{Graduate Student Member,~IEEE,} Athanasios P. Chrysologou,\\~\IEEEmembership{Graduate Student Member,~IEEE,} Alexandros-Apostolos A. Boulogeorgos,~\IEEEmembership{Senior Member,~IEEE,} \\  Nestor D. Chatzidiamantis,~\IEEEmembership{Member,~IEEE,} and Harald Haas,~\IEEEmembership{Fellow,~IEEE.}  

\thanks{G. D. Chondrogiannis, A. P. Chrysologou, N. D. Chatzidiamantis and are with the Department of Electrical and Computer Engineering, Aristotle University of Thessaloniki, 54124 Thessaloniki, Greece (e-mails:
\{gchondro,  chrysolog, nestoras\}@auth.gr).} 
\thanks{A-A. A. Boulogeorgos is with the Department of Electrical and Computer Engineering, University of Western Macedonia, 50100 Kozani, Greece (e-mail: al.boulogeorgos@ieee.org).}
\thanks{Harald Haas is with the Department of Engineering, Electrical Engineering Division, Cambridge University, CB3 0FA Cambridge, U.K. (e-mail:
huh21@cam.ac.uk).}
\thanks{This work is supported by the research project MINOAS. The research
project MINOAS is implemented in the framework of H.F.R.I call “Basic
research Financing (Horizontal support of all Sciences)” under the National
Recovery and Resilience Plan “Greece 2.0” funded by the European Union –
NextGenerationEU (H.F.R.I. Project Number: 15857).”
}

}

\vspace{-0.3in}
\maketitle	

\begin{abstract}

In this paper, we identify optical reconfigurable intelligent surfaces (ORISs) as key enablers of next-generation free-space optical (FSO) multiple access systems. By leveraging their beam steering and beam splitting capabilities, ORISs are able to effectively address line-of-sight (LoS) constraints, while enabling multi-user connectivity. We consider an ORIS-assisted non-orthogonal multiple access (NOMA) system model consisting of a single transmitter (Tx) and two receivers (Rxs). We derive novel analytical expressions to characterize the statistical particularities of the Tx-ORIS-Rx communication channel. Building upon the aforementioned expressions, we investigate the outage performance of the Rxs by deriving exact analytical expressions for the outage probability (OP) of each Rx. To provide deeper insights into the impact of various system parameters and physical conditions on the outage performance of each Rx, we conduct a high signal-to-noise ratio (SNR) analysis, that returns asymptotic expressions for the Rxs' OPs at the high-SNR regime. Monte Carlo simulations validate the analysis, demonstrate the effectiveness of ORIS-enabled NOMA under a variety of configurations and physical scenarios, and showcase its superiority over its orthogonal-based counterpart.
\end{abstract}


\section{Introduction} 

\IEEEPARstart{F}{ree-space} optical (FSO) communication has garnered significant attention in recent years due to its compelling advantages and potential applications. FSO has been thoroughly explored in research for its ability to deliver several key benefits, including low error rates, unlicensed operation, and enhanced security. FSO communication is particularly appealing because it offers a cost-effective solution for establishing high-bandwidth links over relatively short distances \cite{FSO_theory}.
The versatility of FSO technology extends across various domains, making it suitable for both terrestrial and extraterrestrial communication scenarios. Given its potential to address growing demands for high-speed data transmission and its suitability for challenging environments, FSO represents a transformative approach to next-generation communication networks \cite{FSO_theory_2, Haas_Survey}. Some indicative examples of the application of FSO include robust connections between buildings, smart city networks for internet-of-things (IoT) and traffic systems, temporary event connectivity, fifth-generation (5G) backhaul in urban areas, and links between satellites and ground stations \cite{FSO_app}. Nevertheless, a significant hurdle in the development of FSO systems lies in ensuring direct line-of-sight (LoS) connections between optical transmitters (Tx) and receivers (Rxs), which restricts their applicability.

To address this challenge, a cutting-edge solution known as optical reconfigurable intelligent surfaces (ORIS) has been presented (see \cite{nature_1}, \cite{nature_2} and reference therein). ORIS technology enables the creation of virtual line-of-sight (LoS) connections, effectively bypassing environmental barriers that would otherwise obstruct the optical path. By strategically positioning ORIS in obstructed pathways, it can precisely redirect optical signals, enabling reliable data transmission even in complex environments. Unlike relays, ORIS consists of passive elements that can be seamlessly integrated into existing infrastructure, offering a cost-effective solution \cite{schober_survey}. 

ORIS are generally classified into two types: (i) mirror array-based, and (ii) optical metasurface-type ORISs. The mirror-array-based ORIS incorporates a two-dimensional array of deflectable micro-mirror units capable of optimizing system performance. By mechanically adjusting the orientation of the mirrors within the array, it enables precise control of signal reflection angles, effectively directing optical signals toward the Rx \cite{Magaz,schober_survey}.
In contrast, metasurface-based ORISs utilize advanced nanotechnology and materials science to create reflective surfaces capable of beam-steering and beamforming \cite{Magaz}. These surfaces mainly consist of flat arrays of sub-wavelength resonant elements that modulate the phase of incoming beams, enabling precise control over beam directionality, such as achieving anomalous reflection in accordance with the generalized law of reflection \cite{rec_beam, Optical_adaptive, optical_met}. Besides, mechanical steering leads into long response time; thus, their suitability in real-world FSO applications, in which the coherence time is in the order of hundreds $\mu s$ is questionable, The metasurface-based ORIS response time depends on the phase-switching mechanism (thermal, electro-thermal, or electrical) and is usually lower than $1 \rm{\mu s}$. This is the main reason for which the photonics society steered their attention towards metasurface-based ORIS. 

Understanding ORIS's fundamental properties is essential for deploying the corresponding optical wireless systems. Accurate models are vital for predicting performance, and addressing practical challenges in real-world applications. In preliminary modeling efforts of ORIS-assisted optical wireless systems, a key challenge is determining whether to portray the surface as a continuous entity or as an array of discrete elements. Unlike their counterparts in the RF bands, ORIS can be effectively modeled as a single mirror \cite{roadmap}. Owing to the high directivity of optical signals, a single mirror is sufficient to achieve beam reflection and deflection. Another critical task is accurately modeling the Rx’s pointing errors, which occur as the optical beam traverses the ORIS-generated path. 

Particularly, the works in \cite{Alou_ORIS, Unified} consider a discrete metasurface composed of multiple reflecting elements, each subject to independent channel fading, and employ a statistical model analogous to the zero-boresight pointing error model. In more detail, the study in \cite{Alou_ORIS} introduces compact approximate closed-form expressions for evaluating the performance of ORIS-assisted systems by employing the central limit theorem approximation, while  \cite{Unified} articulated extensive and unified closed-form expressions applicable to both single and cascaded ORIS configurations across a range of fading distribution models. Furthermore, \cite{Array-type} conducted a deterministic analysis of reflected beams, deriving closed-form expressions for the output power density distribution and power efficiency of optical relay systems employing micro-mirror array and metasurface-based array configurations.

In contrast, the analyses in \cite{Dobre_Haas, photonics_1, boulo_fso, Najafi_new, optics_express} adopted a continuous-surface model for the reflecting structure, effectively treating it as an anomalous reflector for the optical beam. Specifically, \cite{Dobre_Haas} was based on a pointing error model similar to the zero-boresight approach, while offering an in-depth analysis of the Rx’s beam characteristics. In parallel, the authors of \cite{photonics_1} employed a zero-boresight model and made a straightforward approximation of the beam's displacement at the Rx. Also, in\cite{boulo_fso}, the overall performance of a cascaded multi-RIS-enabled FSO system is evaluated under the assumption of a zero-boresight model for each cascaded link.  In \cite{Najafi_new}, the authors conducted a detailed statistical analysis of pointing errors at the Rx, modeling the entire communication pathway as a single unified channel coefficient. This approach deviates from traditional methods that treat Tx-to-ORIS and ORIS-to-Rx links as separate entities and also incorporates the effects of building sways, which influences all system nodes. Furthermore, the authors proposed statistical channel models applicable to both two-dimensional (2D) and three-dimensional (3D) cases. Finally, \cite{optics_express} addressed the outage-guaranteed transmission challenges associated with link selection and power configuration by utilizing the 2D pointing error model that was presented in \cite{Najafi_new}.

In the meantime, one of the most critical challenges facing the upcoming 6G networks is the ability to support massive connectivity in high-density scenarios, achieve ultra-high spectral efficiency, and enable dynamic and flexible resource allocation \cite{NOMA_NGMA_1}. Addressing these requirements is essential for realizing the full potential of next-generation FSO networks. So far, orthogonal multiple access (OMA)-based approaches, such as time division multiple access (TDMA), wavelength division multiple access (WDMA), code division multiple access (CDMA) and space division multiple access (SDMA), have been widely adopted as solutions for multiple access in FSO networks \cite{jahid2022contemporary}. These techniques leverage the inherent features of FSO, such as high directionality and spatial isolation, to efficiently allocate resources among Rxs while minimizing interference. However, traditional OMA-based schemes have demonstrated limitations in addressing one of the most critical demands of future 6G communication networks, i.e., enabling massive, limitless, and flexible access \cite{NOMA_NGMA_3, NOMA_NGMA_2, NOMA_NGMA_1}.  Non-Orthogonal Multiple Access (NOMA), on the contrary, allows multiple users to share the same resources by leveraging differences in power levels, thereby significantly enhancing spectral efficiency and user connectivity compared to OMA-based counterparts \cite{NOMA_NGMA_3}. Recognizing these advantages, except from the the realm of radio frequency (RF) communications where NOMA has been extensively studied and is reported as a leading candidate of the so-called next-generation multiple access (NGMA) paradigm \cite{NOMA_NGMA_3, NOMA_NGMA_1}, in the last few years both academia and industry have also turned their attention to the integration of NOMA in millimeter-wave (mmWave), Terahertz (THz) as well as visible light communication (VLC) systems (see \cite{NOMA_NGMA_3,liu2024road,jiang2024terahertz,NOMA_NGMA_1} and references therein), where the results have been highly promising. The demonstrated success of NOMA in these frequency domains motivates the investigation of its potential application in FSO communication systems. However, the application of NOMA to FSO systems is not without challenges, primarly due to the highly directional nature of FSO beams, which contradicts the overlapping coverage principle of NOMA.  In this context, a limited number of studies have explored the integration of NOMA into FSO systems, focusing primarily on uplink scenarios and their performance analysis \cite{Diamantoulakis,noma_arxiv}, or experimental implementations \cite{experimental}. Considering that ORISs are capable of facilitating beam focusing, beam splitting, and multi-beam power allocation \cite{Magaz}, a pertinent question arises: can ORISs support the implementation of multiple access in FSO systems? This study addresses this question by proposing the utilization of ORISs as a \emph{multiple access enabler} for next-generation FSO networks. By leveraging their beam steering and beam splitting capabilities, ORISs can dynamically manage beam control and power distribution among multiple users, effectively realizing NOMA at the physical layer. This approach directly addresses the scalability and connectivity demands of future FSO systems, making ORISs a transformative solution for enabling advanced multiple access in next-generation networks.

Motivated by the aforementioned considerations, this work investigates the performance of the ORIS-enabled NOMA system for FSO communication, between a transmitter (Tx) and two designated Rxs, namely Rx1 and Rx2. By employing the meticulous statistical pointing errors 3D model introduced in \cite{Najafi_new}, we derive analytical expressions to assess the outage performance of the proposed system. To the best of our knowledge, this work presents the first attempt to incorporate ORIS technology into a downlink multiple-access FSO system designed to serve two specific Rxs. 
The simulation results, in agreement with the theoretical analysis, demonstrate the effectiveness of NOMA in potential scenarios, highlighting its suitability as a multiple access technique compared to OMA.

In particular, the technical contribution of this paper is summarized as follows: 
\begin{itemize}
    \item Adopting the novel 3D geometric and misalignment loss (GML) model of \cite{Najafi_new}, we derive analytic expressions for the probability density function (PDF), cumulative distribution function (CDF), and outage probability (OP) of the end-to-end (E2E)  ORIS-assisted communication system.
    \item We apply the aforementioned analysis to derive analytic expressions for the OP of the two Rxs in the proposed ORIS-enabled NOMA system between a Tx and Rx1, Rx2.
    \item Additionally, we perform a high signal-to-noise ratio (SNR) analysis of the proposed system which returns insightful asymptotic expressions for the OP of the two Rxs.
    \item Finally, simulations are performed to validate the derived expressions and to examine the impact of various system parameters, including the transmit SNR, the power allocation coefficients for the intended user signals, the power of the reflected beams, the variance of fluctuations caused by building sway, and the overall link distances, under a range of turbulence conditions. Moreover, the simulation results and theoretical analysis consistently demonstrate the effectiveness of NOMA across a range of scenarios compared to OMA.
\end{itemize}

The remainder of this paper is organized as follows: The proposed system model as well as the signal and channel models are presented in Section \ref{sysmdl}. Section III provides the statistical characterization of the adopted channel models and investigates the OP performance of the Rxs in the presented system setup. In Section IV, simulations and analytical results verify the correctness of the provided mathematical framework, while, in Section V, we summarize the main findings of this contribution and highlight the paper's message.

\emph{Notations}: The absolute value, exponential, and natural logarithm functions are, respectively, denoted by
$|{\cdot}|, \exp{(\cdot)}, \text{
and } \ln (x)$. $\mathbb{P}(\mathcal{A})$ denotes the probability for the event $\mathcal{A}$ to be valid. The operators $\max{(\cdot)}$ and $\min{(\cdot)}$ represent the maximum and minimum. Also, $
\left\{p_i\right\}_1^N=\left\{p_1, \cdots, p_N\right\}$ is the shorthand notation for a finite sequence of $N$ elements. Moreover, $I_0(x), K_\nu(x)$ and $\Gamma(x)$ represent the zero-th order modified Bessel function of the first kind \cite[eq. (8.447/1)]{Gradshteyn2014}, the $\nu$-th order modified Bessel function of the second kind \cite[eq. (8.432/1)]{Gradshteyn2014}, and the Gammma function \cite[eq. (8.310/1)]{Gradshteyn2014}, respectively. Finally, $G_{p, q}^{m, n}\left(x\, \Big{|} \begin{array}{c}a_1, a_2, \cdots, a_p \\ b_1, b_2, \cdots, b_q\end{array}\!\!\right)$ stands for the Meijer G-function \cite[eq.(9.301)]{Gradshteyn2014}.

\section{System model}\label{sysmdl}

This section is focused on presenting the exact modeling of the proposed ORIS-enabled system. We firstly introduce the system and signal models, detailing the operation principles of the ORIS. Then, the statistical models adopted to describe the channels' particularities are presented.

\subsection{System and signal models}

\begin{figure}[t]
\centering
	\includegraphics[scale = 0.85]{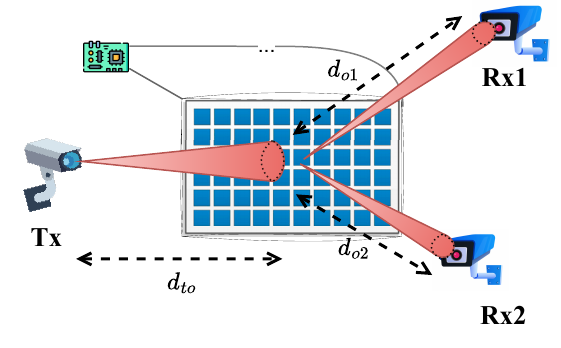}	
 \caption{Schematic representation of the ORIS's beam-splitting function and the equivalent E2E distances.}
	\label{Fig:sys_model}
\end{figure}

As depicted in Fig. \ref{Fig:sys_model}, we consider an ORIS-assisted intensity modulation and direct detection (IM/DD) FSO system model, where a Tx intents to serve two Rxs. We consider that Tx encodes data by adjusting the intensity of a laser beam, while at the Rxs side, a lens concentrates the incoming optical power onto a photodetector (PD), which then gauges the intensity of the received optical signal. It is assumed that, without loss of generality, the distance between ORIS-Rx1 is greater than the distance of ORIS-Rx2 and that no-direct links can be established between, Tx and Rx1, Rx2; thus, the communication between Tx and Rx1, Rx2 takes place only via the ORIS. Furthermore, it is considered that NOMA is exploited for the Rxs' access in order to allow maximization of the bandwidth utilization. Consequently, two seperate and independent communication links are realized through the ORIS that maintain a LOS with both the Tx and Rx1, Rx2. 

As illustrated in Fig. 1, the aperture of the laser source (LS) aims at the ORIS, which splits the incoming optical beam toward the two Rxs' lenses. The lens then focuses the two splitted beams onto the PD of each Rx to capture the optical power. It is also assumed that the ORIS unit has the capability of splitting the incident beam into two separate reflective beams with equivalent phase-shift profiles and different beam widths, \cite{Array-type, Najafi_new}. 

According to the NOMA principle, the superimposed transmitted optical  signal $x\in \mathbb{R}^+$ by Tx can be expressed as

\begin{equation}
    x=\sqrt{a_1 P} x_1+\sqrt{a_2 P} x_2, 
\end{equation}
where $x_j$, with $j \in \{1,2\}$, is the  signal intended for Rx$j$, $P$ is Tx's transmission power and $a_1, a_2$ are the power allocation coefficients with $a_1>a_2$ and $a_1+a_2=1$. By accounting for the ORIS beam splitting functionality, the received signals at Rx1 and Rx2 can be modeled as follows:
\begin{equation}
    y_1=B_1 \, h_1 \, x \, + \, n_1
\end{equation}
and
\begin{equation}
    y_2=B_2 \, h_2 \, x \,+ \,n_2,
\end{equation}
where $n_1$ $\in \mathbb{R}$ and $n_2$ $\in \mathbb{R}$ are the independent zero-mean, real-valued additive Gaussian noises with variance $\sigma_{n}^2$ at Rx1 and Rx2. 
Additionally, $B_1$ and $B_2$ denote the beam splitting factors, i.e., $B_1$, $B_2$ are parameters that quantify the proportion of power allocated to each of the two reflected beams emerged from the beam-splitting operation. We also assumed a normalized unitary Rx's responsivity, which is not a limiting assumption, as the results can be readily generalized to appropriate values. Finally, $h_j$ $\in \mathbb{R}^+$ denotes the E2E channel coefficient between the Tx and RX$j$.

Regarding the decoding process at the Rxs' side, Rx1 directly decodes $x_1$, treating the interference caused by $x_2$ as noise; thus, the instantaneous electrical signal-to-interference-plus-noise ration (SINR) for the decoding of $x_1$ can be expressed~as
\begin{align}
	\gamma_{1} = \frac{ a_1 \, B_1 \, P \, h_1^2}{ a_2 \, B_1 \, P \, h_1^2 \, + \, \sigma_{n}^2} \, = \, \frac{ a_1 \, B_1 \, \, \Bar{\gamma} \, h_1^2}{ a_2\, B_1 \, \Bar{\gamma} \, h_1^2  \, + \, 1}, \label{g1}
\end{align}
where $\Bar{\gamma} = \frac{P}{\sigma_{n}^2}$ denotes the transmit SNR.
On the other hand, Rx2 firstly decodes $x_1$ and then applies SIC in order to decode its intended signal $x_2$. Hence, the instantaneous SINRs for the decoding of $x_1$ and $x_2$ at Rx2 can be respectively written~as 
\begin{align}
	\gamma_{21}=\frac{a_1 \, B_2 \, \Bar{\gamma} \, h_2^2}{a_2 \, B_2 \, \Bar{\gamma} \, h_2^2 \, + \, 1},  \label{g21}
\end{align}
and
\begin{align}
	\gamma_{22} = a_2 \, B_2 \, \Bar{\gamma} \, h_2^2 . \label{g22}
\end{align}

\subsection{Channel and statistical models}

The E2E channel coefficient $h_j$ for the Rx$j$, with $j = 1,2$, is considered to be the product of three factors \cite{Hralinovic,Najafi_new}: 
\begin{equation}
    h_j = h_{lj} h_s h_{gj}.
\end{equation}
Here, $h_{lj}$ denotes the atmospheric loss at $\mathrm{Rx}_j$, $h_{gj}$ captures the pointing errors at $\mathrm{Rx}_j$, and $h_s$ models the atmospheric turbulence, which is assumed to be the same for both Rxs.
To simplify notation initially, we will omit the use of the pointer $j$, which corresponds to the number of the Rx, i.e., $h = h_l h_s h_g$. In what follows, we present the models adopted for $h_l$, $h_s$ and $h_g$.

\subsubsection{Atmospheric loss}
Practical reflecting surfaces can also absorb or scatter part of the beam's power, with this behavior being affected by factors such as the operating frequency and the bias voltage applied to the ORIS surface. The atmospheric loss factor is given by:
\begin{equation}
h_l=\rho 10^{-\sigma d_{z} / 10},
\end{equation}
where $\rho$ is the reflection efficiency, $\sigma$ the attenuation coefficient and $d_{z}$
represents the E2E transmission distance of the optical beam, defined as the sum of the distances from the Tx to the ORIS, $d_{to}$, and the distance from the ORIS to the Rx$j$, $d_{oj}$, as depicted in Fig. 1. For FSO communication systems operating at a wavelength of $\lambda = 1550 \, \text{nm}$, typical values for $\rho$ range between 0.7 and 1 \cite{Najafi_new}, and the values of $\sigma$ are determined by the prevailing weather conditions, as described by the Kim's model in \cite{Kim}.

\subsubsection{Pointing errors}
\begin{figure}[t]
\centering	\includegraphics[scale = 1.30]{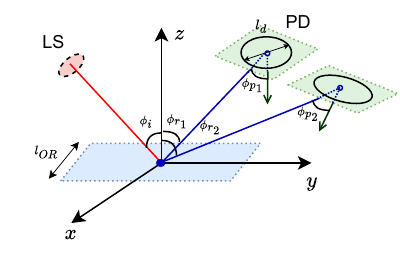}	
 \caption{Schematic representation of the considered system's geometrical characteristics.}
	\label{Fig:geometry}
\end{figure}

In an ORIS-enabled link, the laser beam targets the ORIS instead of the Rx's lens. However, fluctuations in the beam’s footprint center on the ORIS due to pointing errors cause misalignment at the Rx plane. The impact of these errors depends on the ORIS's phase-shift design or orientation \cite{Jamali}.
Assuming Gaussian fluctuations for the building
sways of the Tx, ORIS, and Rx, the PDF of $h_g$ can be expressed as \cite{Najafi_conf}:
\begin{equation}
f_{h_g}(h_g)\!=\!\frac{\omega}{A_0}\!\left(\frac{h_g}{A_0}\right)^{\frac{\left(1+q^2\right) \omega}{2 q}-1}\!\!I_0\left(\!-\frac{\left(1-q^2\right) \omega}{2 q} \ln \left(\frac{h_g}{A_0}\right)\!\right),
\label{f_hg}
\end{equation}
where $0 \leq h_g \leq A_0$, with $A_0$ denoting the fraction of collected power at the center of PD. The parameters $\omega$ and $q$  are given by $\omega=\frac{\left(1+q^2\right) t w^2\left(d_{z}\right)}{4 q \Omega}$ and $q=\frac{\sigma_{u_2}}{\sigma_{u_1}}$, where $w\left(d_{z}\right)$ is the beam width
at the Rx of distance $d_z$, and $\Omega=\sigma_{u_1}^2+\sigma_{u_2}^2$, with $\sigma_{u_1}^2$, $\sigma_{u_2}^2$ being the variances of the Gaussian misalignment vector on the PD plane and are given by \cite{Najafi_conf}
\begin{equation}
\sigma_{u_1}^2=\frac{1}{\sin ^2\left(\phi_p\right)}\left(\sigma_s^2+4 \cos ^2\left(\phi_r\right) \sigma_r^2+\sigma_p^2\right)  
\end{equation}
and 
\begin{equation}
\sigma_{u_2}^2=\frac{1}{\sin ^2\left(\phi_p\right)}\left(\sigma_s^2+\sigma_p^2\right),
\end{equation}
respectively. The variances $\sigma_s$, $\sigma_r$ and $\sigma_p$ denote the building sways for the Tx, the ORIS and the Rx, respectively.  Furthermore, $t=\sqrt{t_1 t_2}$, where $t_1=\frac{\sqrt{\pi} \operatorname{erf}\left(\nu_1\right)}{2 \nu_1 \exp \left(-\nu_1^2\right)}$, $ t_2=\frac{\sqrt{\pi} \operatorname{erf}\left(\nu_2\right)}{2 \nu_2 \exp \left(-\nu_2^2\right) \sin ^2\left(\phi_p\right)}$, $\nu_1=\frac{l_{d}}{2w\left(d_{z}\right)} \sqrt{\frac{\pi}{2}}$, and $\nu_2=\nu_1\left|\sin \left(\phi_p\right)\right|$. 
As shown in Fig. \ref{Fig:geometry}, the parameters, $l_{d}$, $\phi_p$ and $\phi_r$ represent, respectively, the length of the PD, the angle between the reflected beam and the PD, and the angle between the laser beam and the ORIS plane. It is important to highlight that, for the sake of simplifying the system parameters, we have adopted the assumption that the ORIS functions as a mirror with mechanical rotation, such that $\phi_{i} = \phi_{r}$. However, this assumption is not restrictive, as it can be demonstrated that for a given phase-shift profile and a metasurface-based ORIS, there exists an equivalent mirror-assisted (with equivalent beam width) capable of producing a reflected electric field on the mirror that is identical to the one generated on the ORIS in the original configuration \cite{Najafi_new}. In the context of evaluating the overall performance of ORIS-enabled communication, this assumption does not significantly affect the analysis and can be readily generalized to the exact scenario. For additional details regarding the assumptions of the geometrical model and its parameters, please refer to \cite{Najafi_conf,Najafi_new}.

\subsubsection{Atmospheric turbulence}
Atmospheric turbulence induces the scintillation effect, leading to signal fluctuations at the Rx. The Gamma-Gamma (GG) model is widely regarded as one of the most appropriate statistical models for characterizing a broad spectrum of turbulence conditions, ranging from weak to strong. Its PDF is given as in \cite{GG}:
\begin{equation}
 f_{h_s}(h_s)=\frac{2(\alpha \beta)^{\frac{\alpha+\beta}{2}}}{\Gamma(\alpha) \Gamma(\beta)} {h_s}^{\frac{\alpha+\beta}{2}-1} K_{\alpha-\beta}\left(2 \sqrt{\alpha \beta h_s}\right),
 \label{f_hs}
\end{equation}
where $\alpha$ and $\beta$ are parameters directly related to the effects induced by the large-scale and small-scale scattering, respectively. These parameters are given by 
\begin{align}
\alpha & =\left[\exp \left(\frac{0.49 \sigma_R^2}{\left(1+1.11 \sigma_R^{12 / 5}\right)^{7 / 6}}\right)-1\right]^{-1} \nonumber \\
\beta & =\left[\exp \left(\frac{0.51 \sigma_R^2}{\left(1+0.69 \sigma_R^{12 / 5}\right)^{5 / 6}}\right)-1\right]^{-1}
\end{align}
where $\sigma_{R}^2 = 1.23 C_n^2 k^{7 / 6} d_{z}^{11 / 6}$ is the Rytov variance and $k=2 \pi / \lambda$ is the wavenumber, in which $\lambda$ is the wavelength, $d_z$ is the E2E channel distance, and $C_n^2$ is the index of refraction structure parameter. In general, $C_n^2$ varies from $10^{-13} \mathrm{~m}^{-2 / 3}$ to $10^{-17} \mathrm{~m}^{-2 / 3}$. Typically, weak turbulence fluctuations are associated with $\sigma_R^2<1$, moderate with $\sigma_R^2 \approx 1$, and strong with $\sigma_R^2>1$ \cite{turbulence, Karagiannidis}.




\section{Performance Analysis}
In communication scenarios where maintaining high quality of service (QoS) experience is critical for Rxs, OP serves as a particularly meaningful performance evaluation criterion. In this context, this section investigates Rxs' outage performance in the presented ORIS-enabled system. Firstly, we derive statistical characterizations of the Tx-ORIS-Rx communication channel by extracting analytical expressions for the E2E channel's PDF and CDF. Subsequently, these expressions are utilized in order to extract analytical expressions for Rxs' OPs in the presented ORIS-enabled NOMA system model. A high SNR analysis is also provided which provides useful insights regarding the outage behavior of the RXs in the high SNR regime.

\subsection{Statistical characterization of the E2E channel}
Towards facilitating performance investigation of ORIS-enabled communication systems, it is essential to derive an expression for the PDF of the E2E Tx-ORIS-Rx channel model. In this direction, the PDF of the E2E channel, which is defined by the product of two random variables, $h_s$ and $h_g$, and scaled by the atmospheric loss term $h_l$, can be derived as follows \cite{Hralinovic}:
\begin{equation}
    f_h\left(h\right) = \int_{\frac{h}{A_0 h_l}}^{\infty} \frac{1}{h_s h_l} f_{h_g}\left(\frac{h}{h_s h_l}\right)f_{h_s}\left(h_s\right)dh_s.
    \label{f_h}
\end{equation}
In the following theorem, we provide an analytical expression for the PDF of the E2E channel.
\begin{theorem}
    The PDF of the E2E channel model
    can be analytically expressed as:
    \begin{align}
	f_{h}&(h) =\frac{\omega\alpha\beta}{h_l A_0 \Gamma(\alpha) \Gamma(\beta)}\sum_{k=0}^{\infty} \binom{2k}{k}\left(\frac{v}{2}\right)^{2k}
	\nonumber\\ & \times\mathrm{ G}_{2k+1,2k+3}^{2k+3,0}\!\left(\!\alpha \beta \frac{h}{A_{0} h_l} \!\! \mathrel{\bigg|} \begin{array}{c}
	    \left\{c\right\}_{1}^{2k+1} \\
	    \!\!\!\! \alpha-1, \beta-1, \left\{c-1\right\}_{1}^{2k+1}\!\!\!\!
	\end{array}\!\right),
    \label{eq:pdf}
\end{align}
where $c = \frac{\left(1+q^2\right) \omega}{2q}$ and $v = \frac{\left(1-q^2\right) \omega}{2q}$.
\end{theorem}
\begin{IEEEproof}
    The proof is provided in Appendix A.
\end{IEEEproof}
Leveraging the PDF expression of (\ref{eq:pdf}), in what follows we provide a proposition that returns an analytical expression for the E2E channel's CDF.
\begin{theorem}
    The CDF of the E2E channel model can be analytically expressed as:
    \begin{align}
	F_{h}(h) &=\frac{\omega}{\Gamma(\alpha) \Gamma(\beta)}\sum_{k=0}^{\infty} \binom{2k}{k}\left(\frac{v}{2}\right)^{2k}
	\nonumber\\ & \times\mathrm{ G}_{2k+2,2k+4}^{2k+3,1}\left(\alpha \beta \frac{h}{A_{0} h_l} \mathrel{\bigg|} \begin{array}{c}
		\!\!1, \left\{c+1\right\}_{1}^{2k+1} \\
	    \!\! \alpha, \beta, \left\{c\right\}_{1}^{2k+1}\!, 0 \!\!
	\end{array}\right).
    \label{eq:cdf}
\end{align}
\end{theorem}
\begin{IEEEproof}
    The CDF is typically derived as $F_{h}(h) = \int_{0}^{h} f_{h}(x)dx$. 
    By firstly utilizing \cite[(26)]{algorithm} in conjunction with \cite[(9.31.5)]{Gradshteyn2014}, and then applying some appropriate algebraic manipulations, \eqref{eq:cdf} can be derived.
\end{IEEEproof}

For the sake of completeness, and for the special case of a single Rx, in the following lemma we provide an analytical expression for the OP of the Rx.
\begin{lemma}
    The OP of a single Rx in a Tx-ORIS-Rx system setup can be derived as:
    \begin{align}
	P_{\text{o}}(\gamma^{\text{th}}) &\!=\!\frac{\omega}{\Gamma(\alpha) \Gamma(\beta)}\sum_{k=0}^{\infty} \binom{2k}{k}\left(\frac{v}{2}\right)^{2k}
	\nonumber\\ & \times\mathrm{ G}_{2k+2,2k+4}^{2k+3,1}\left( \frac{\alpha \beta }{A_{0} h_{l}}\sqrt{\frac{\gamma^{\text{th}}}{\Bar{\gamma}}} \mathrel{\bigg|} \begin{array}{c}
		\!\!1, \left\{c+1\right\}_{1}^{2k+1} \\
	    \!\! \alpha, \beta, \left\{c\right\}_{1}^{2k+1}\!, 0 \!\!
	\end{array}\right),
    \label{out_p}
    \end{align}
    where $\gamma^{\text{th}}$ denotes the SNR threshold for successful decoding of Rx's message.
\end{lemma}
\begin{IEEEproof}
    With respect to the fact that the OP is defined as the probability that the SNR at the Rx's side falls below $\gamma^{\text{th}}$, Rx's OP can be given as 
    \begin{equation}
        P_{\text{o}} = \mathbb{P}\left(\Bar{\gamma} \, h^2 <\gamma^{\text{th}}\right) = F_{h}\left(\sqrt{\frac{\gamma^{\text{th}}}{\Bar{\gamma}}}\right). \label{OP_singleRx}
    \end{equation}
    Invoking (\ref{eq:cdf}) in (\ref{OP_singleRx}), (\ref{out_p}) can be straightforwardly derived.
\end{IEEEproof}

\subsection{NOMA}
Assuming constant RXs' target rates, i.e. $R_{1}$ and $R_{2}$ for Rx1 and Rx2, respectively, then OP is an appropriate evaluation criterion for system's performance. An outage occurs when the instantaneous SINR falls below a predefined threshold $\gamma^{\text{th}}_{j}=2^{R_j}-1$. In what follows, we provide two theorems that return analytical expressions for Rx1's as well as Rx2's OP. 

\begin{proposition}
    The OP of Rx1 can be evaluated as
\begin{equation}
		P_{1}^{o}= \begin{cases}
        F_{h_1}\left(\sqrt{\frac{\gamma^{\text{th}}_{1}}{a_1 B_1 \Bar{\gamma}-a_2 B_1 \Bar{\gamma} \gamma^{\text{th}}_{1}}}\right) , & \frac{a_1}{a_2} > \gamma^{\text{th}}_{1} \\
			1, & \frac{a_1}{a_2} \leq \gamma^{\text{th}}_{1}
		\end{cases},
        \label{Rx1_OP}
\end{equation}
where $F_{h_1}(\cdot)$ denotes the CDF of $h_1$, which is provided by (\ref{eq:cdf}).
\end{proposition}

\begin{IEEEproof}
We note that an outage for Rx1 occurs when Rx1 cannot decode its message. Hence,
\begin{equation}
		P_{1}^{o}=\mathbb{P}(\gamma_{1}<\gamma^{\text{th}}_{1}). 
        \label{U1_out_v1}
\end{equation} 
Utilizing (\ref{g1}), (\ref{U1_out_v1}) can be transformed into
\begin{equation}
	\begin{split}
		P_{1}^{o}&= \mathbb{P}\left(\frac{a_1 \, B_1 \, \Bar{\gamma} \, h_1^2}{a_2 \, B_1 \, \Bar{\gamma} \, h_1^2  \, + \, 1} <\gamma^{\text{th}}_{1}\right) \\
        &= \begin{cases}
			\mathbb{P}\left(h_{1}< \sqrt{\frac{\gamma^{\text{th}}_{1}}{a_1 B_1 \Bar{\gamma}-a_2 B_1 \Bar{\gamma} \gamma^{\text{th}}_{1}}}\right) , & \frac{a_1}{a_2} > \gamma^{\text{th}}_{1} \\
			1, & \frac{a_1}{a_2} \leq \gamma^{\text{th}}_{1}
		\end{cases}.  
	\end{split} \label{U1_out_v2}
\end{equation}
   Leveraging the CDF of $h_1$, we conclude to (\ref{Rx1_OP}).
\end{IEEEproof}

\begin{remark} \label{rem_Rx1}
    From (\ref{Rx1_OP}), it occurs that $\frac{a_1}{a_2} > \gamma^{\text{th}}_{1}$ serves as an `operation' condition for Rx1, i.e, if this condition does not hold then Rx1 lies in a constant outage.
\end{remark}
    
Compared to the case of Rx1 which directly decodes its intented message, in order to obtain its message, Rx2 should exploit the SIC principle. In the following theorem, the OP of Rx2 is provided. 
\begin{proposition}
  The OP of Rx2 in the proposed ORIS-enabled system model can be evaluated as   
  \begin{equation}
    \begin{split}
		P_{2}^{o}=\!\begin{cases}
			F_{h_2}\left(\sqrt{\max \left\{ \frac{\gamma^{\text{th}}_{1}}{a_1 B_2 \Bar{\gamma}-a_2 B_2 \Bar{\gamma} \gamma^{\text{th}}_{1}} , \frac{\gamma^{\text{th}}_{2}}{a_2 B_2 \Bar{\gamma} } \right\}}\right)\!, & \frac{a_1}{a_2} > \gamma^{\text{th}}_{1} \\
			1, & \frac{a_1}{a_2} \leq \gamma^{\text{th}}_{1}\!,
		\end{cases} 
	\end{split} \label{Rx2_OP}  
  \end{equation}
  where $F_{h_2}(\cdot)$ denotes the CDF of $h_2$ which has been extracted in an analytical closed-form in (\ref{eq:cdf}).
\end{proposition}

\begin{IEEEproof}
Accounting that at the Rx2 side the decoding order is $(x_{1},x_{2})$, in order to avoid outage Rx2 must firstly successfully decode $x_{1}$ and then its own message $x_{2}$ exploiting SIC. Thus,
\begin{equation}
		P_{2}^{o}=1-\mathbb{P}(\gamma_{21}\geq \gamma^{\text{th}}_{1},\gamma_{22}\geq \gamma^{\text{th}}_{2}) \label{U2_out}
\end{equation}   
Invoking (\ref{g21}) and (\ref{g22}) in (\ref{U2_out}), yields
\begin{equation}
	\begin{split}
		P_{2}^{o}&=1-\mathbb{P}\!\left(\!\frac{a_1 \, B_2 \, \Bar{\gamma} \, h_2^2}{a_2 \, B_2 \, \Bar{\gamma} \, h_2^2 \, + \, 1}\!\geq\! \gamma^{\text{th}}_{1}, \, a_2 \, B_2 \, \Bar{\gamma} \, h_2^2 
        \!\geq\! \gamma^{\text{th}}_{2} \!\right) \\
        &= \!\begin{cases}
			\mathbb{P}\!\left(\!h_{2}\!<\!\sqrt{\max \left\{ \frac{\gamma^{\text{th}}_{1}}{a_1 B_2 \Bar{\gamma}-a_2 B_2 \Bar{\gamma} \gamma^{\text{th}}_{1}} , \frac{\gamma^{\text{th}}_{2}}{a_2 B_2 \Bar{\gamma} } \right\}}\!\right)\!, & \frac{a_1}{a_2}\!>\! \gamma^{\text{th}}_{1} \\
			1, & \frac{a_1}{a_2} \!\leq\! \gamma^{\text{th}}_{1}
		\end{cases} 
	\end{split} \label{U2_out_v2}
\end{equation} 
Utilizing $F_{h_2}(\cdot)$, (\ref{U2_out_v2}) can be rewritten as in (\ref{Rx2_OP}). This completes the proof. 
\end{IEEEproof}

\begin{remark} \label{rem_Rx2}
    Similarly with Rx1, $\frac{a_1}{a_2} \!>\! \gamma^{\text{th}}_{1}$ serves as an `operation' condition for Rx2. This highlights the importance of SIC process at Rx2, since when the aforementioned condition does not hold, Rx2 is unable to decode message $x_1$, and thus, cannot proceed into the decoding of its own message $x_2$, resulting in outage. 
\end{remark}

\subsection{High-SNR analysis} \label{sec:highSNR}
In this section, we provide an asymptotic analysis which provides useful insights regarding Rx1's and Rx2's outage behavior in the high SNR regime. Towards this direction, we initially derive an asymptotic CDF expression for the E2E channel as the argument $h$ approaches zero, i.e., $h\! \rightarrow \!0$. This expression is subsequently utilized to characterize the high-SNR performance for Rx1 and Rx2.

\begin{proposition}
 An asymptotic expression for the CDF of the E2E Tx-ORIS-Rx channel, when $h\! \rightarrow \!0$, can be derived as in \eqref{asympt}, given at the top of the next page.    
 \begin{figure*}
     \begin{equation}
 F_h\left(h\right) \!\stackrel{h \rightarrow 0}{\cong}\!
\begin{cases}
    \displaystyle \frac{\omega\, \Gamma\left(|\beta-\alpha|\right)}{\sqrt{\left(c-\min\left(\alpha,\beta\right)\right)^2-v^2} \, \Gamma\left(\min\left(\alpha,\beta\right)+1\right) \Gamma(\max\left(\alpha,\beta\right))} \displaystyle   \left(\alpha \beta \frac{h}{A_{0} h_l}\right)^{\min\left(\alpha,\beta\right)}, \: \text{for} \:  \min\left(\alpha,\beta\right)<c
    \\
    \\
    \displaystyle \frac{\omega \, \Gamma\left(\alpha-c\right)\Gamma\left(\beta-c\right)}{c\, \Gamma\left(\alpha\right) \Gamma\left(\beta\right)}\left[1+\sum_{k=1}^{\infty} \displaystyle \frac{\left(2k+1\right)^{2k}\left(\frac{v}{2}\right)^{2k}}{\left(2k\right)!^2 \left(k!\right)^2}\left(\ln^{2k} \left(\alpha \beta \frac{h}{A_{0} h_l}\right)\right) \right] \left(\alpha \beta \frac{h}{A_{0} h_l}\right)^c, \: \text{for} \: c< \min\left(\alpha,\beta\right)
\end{cases}
\label{asympt}
\end{equation}
\hrulefill
 \end{figure*}
\end{proposition}

\begin{IEEEproof}
    The proof is provided in Appendix B. Additionally, a convergence analysis of the high-SNR asymptotic expressions is provided in Appendix C.
\end{IEEEproof}
Utilizing the extracted asymptotic expression in \eqref{asympt}, it becomes feasible to forsign the outage behavior of both Rx1 and Rx2 at high transmit SNR values for the proposed ORIS-enabed system setup. Specifically, assuming $\Bar{\gamma} \rightarrow \infty$, then  asymptotic expressions for the OPs of Rx1 and Rx2 can be directly obtained by substituting \eqref{asympt} into \eqref{Rx1_OP} and \eqref{Rx2_OP}, with the appropriate arguments given in these expressions. 

\begin{remark} \label{rem_3}
  By applying \eqref{asympt} to derive the asymptotic OP for Rx1 and Rx2, the resulting diversity order, given by $D = -\lim_{\bar{\gamma} \rightarrow \infty} \frac{\log{P_{1,2}^o}}{\log{\bar{\gamma}}}$, is readily derived as $\min\left(\frac{a}{2}, \frac{\beta}{2}, \frac{c}{2}\right)$. This straightforwardly occurs from the two branches of \eqref{asympt}, considering also the fact that polynomial decay of $\bar{\gamma}^{-c/2}$ dominates the logarithmic growth in the asymptotic limit. Therefore, the multiplicative sum of logarithmic terms modifies the prefactor but does not change the fundamental rate at which the OP goes to zero as $\bar{\gamma} \rightarrow \infty$.
\end{remark}

\begin{remark}
    Based on Remark \ref{rem_3}., the primary factor that influences the asymptotic decay of OP of the Rx1 and Rx2 is the turbulence conditions when $\min\left(\alpha,\beta\right) < c$, whereas the overall pointing errors become the dominant factor affecting performance when $c< \min\left(\alpha,\beta\right)$.
\end{remark}

\section{Numerical Results}\label{numr}
In this section, we analyze the performance of the ORIS-enabled system model by presenting both analytical (an.) and simulation (sim.) results. Unless indicated otherwise, the parameter values adopted for the extraction of the following figures can be found at Table \ref{table:1}. Furthermore, from a computational standpoint, the infinite sum in \eqref{out_p} exhibits rapid convergence. Consequently, in numerical results, the analytical expressions matched the simulation results even when only three terms were included. However, to achieve higher arithmetic precision, the numerical results were obtained by considering $N=10$ terms. Simulations were conducted in MATLAB$\copyright$ and also the Symbolic Math Toolbox was utilized to prevent overflow issues\footnote{Specifically, the \emph{vpa($\cdot$)} function can be utilized to leverage higher-precision arithmetic and thereby prevent floating-point overflows.}.  


\begin{table}[h!]
\centering
\caption{Simulation parameters.}
\resizebox{8.2cm}{!}{%
 \begin{tabular}{|c|c|c|c}
\hline Parameter  & Symbol & Value  \\
\hline 
\hline wavelength  & $\lambda$  & $1550 \text{ nm}$ \\
\hline attenuation coefficient & $\sigma$  & $0.43\times 10^{-3} \text{m}^{-1}$  \\
\hline reflection efficiency & $\rho$  &  0.8\\
\hline lens diameter & $\l_{d}$  & $5\text{ cm}$ \\
\hline Tx-0RIS-Rx$1$ distance & $d_{z1}$  & $1000\text{ m}$ \\
\hline Tx-0RIS-Rx$2$ distance & $d_{z2}$  & $800\text{ m}$ \\
\hline angle b/w PD and reflected beams  & $(\phi_{p1},\phi_{p2})$  & $(\pi/3,\pi/3)$ rad \\
\hline angle b/w ORIS and incident beams  & $(\phi_{r1},\phi_{r2})$  & $(\pi/6, \pi/4)$ rad \\
\hline power allocations & $(a_1,a_2)$ & $(0.9,0.1)$ \\
\hline beam splitting factors & $(G_1,G_2)$ & $(0.4,0.6)$ \\
\hline data rate thresholds & $(R_1^{\text{th}},R_2^{\text{th}})$ & $(2,4.5)$ bit/s/Hz \\
\hline
beam width at the Rxs & $(w(d_{z1}),w(d_{z2}))$ & $(0.45,0.35)$ cm \\
\hline building sways at Tx, ORIS, Rx1, Rx2 & $\sigma_s, \sigma_r, \sigma_l$ &  $0.375 l_d$ cm \\
\hline index of refraction structure & $C_n^2$ & $5\times 10^{-14}$ $\text{ m}^{2/3}$ \\
\hline
\end{tabular}}
\label{table:1}
\end{table}

\begin{figure}[h]
\centering
\includegraphics[width=0.95\columnwidth]{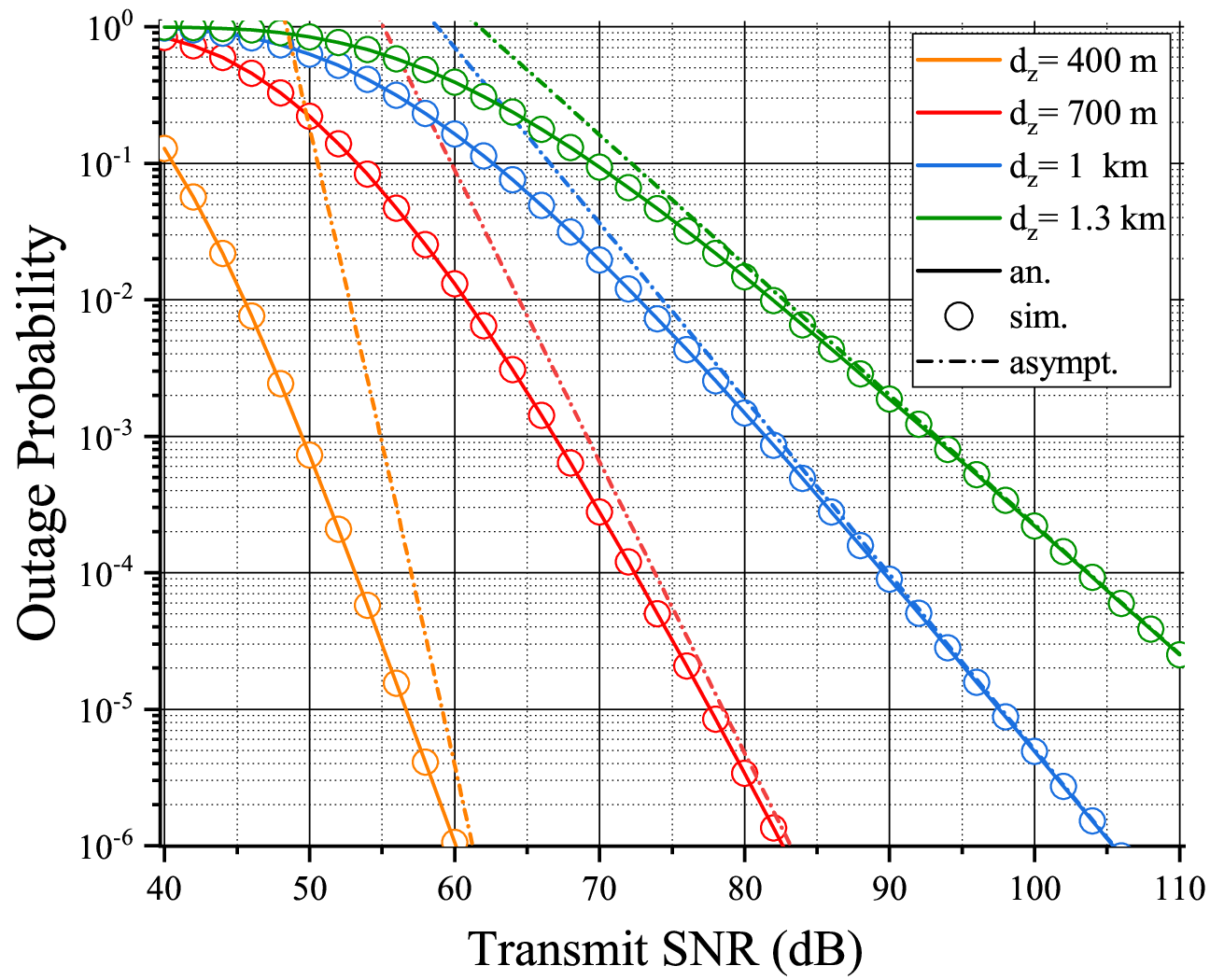}
\caption{OP vs transmit SNR for a single Rx scenario under different E2E Tx-ORIS-Rx distances.}
\label{OP_dz}
\end{figure}

In Fig. \ref{OP_dz}, assuming a single Rx scenario, we present Rx's OP with respect to transmit SNR under different E2E Tx-ORIS-Rx distances and equivalent beam widths at the receiver\cite[(5)]{Najafi_conf}, when the links are subjected to clear weather conditions, laser beam waist radius is 1 mm, and Rx's target data rate equals 1 bit/s/Hz. It becomes obvious that simulations coincide with the analytical results; thus, verifying the accuracy of the analysis and specifically the validity of the extracted E2E channel's CDF in (\ref{eq:cdf}). It is illustrated that the E2E Tx-ORIS-Rx distance $d_z$ has a significant impact on Rx's OP performance since, for a fixed transmit SNR value, as $d_z$ decreases, then Rx's OP substantially decreases. This result was expected, as an increase in E2E distance amplifies the effects of turbulence and atmospheric attenuation while also causing the beam to expand at the RX plane, which in turn reduces pointing errors. Furthermore, the asymptotic expressions for RX's OP are also presented. These expressions closely match the outage probability (OP) curves at high SNR values, thereby validating the asymptotic analysis presented in Section \ref{sec:highSNR} and highlighting the dominant term affecting outage performance. Specifically, for short propagation distances when $c< \min\left(\alpha,\beta\right)$, e.g. $d_z = 400$, pointing errors are the primary limiting factor. However, as the propagation distance increases, the effects of atmospheric turbulence become more pronounced, eventually becoming the dominant factor governing outage probability.

\begin{figure}[h]
\centering
\includegraphics[width=0.95\columnwidth]{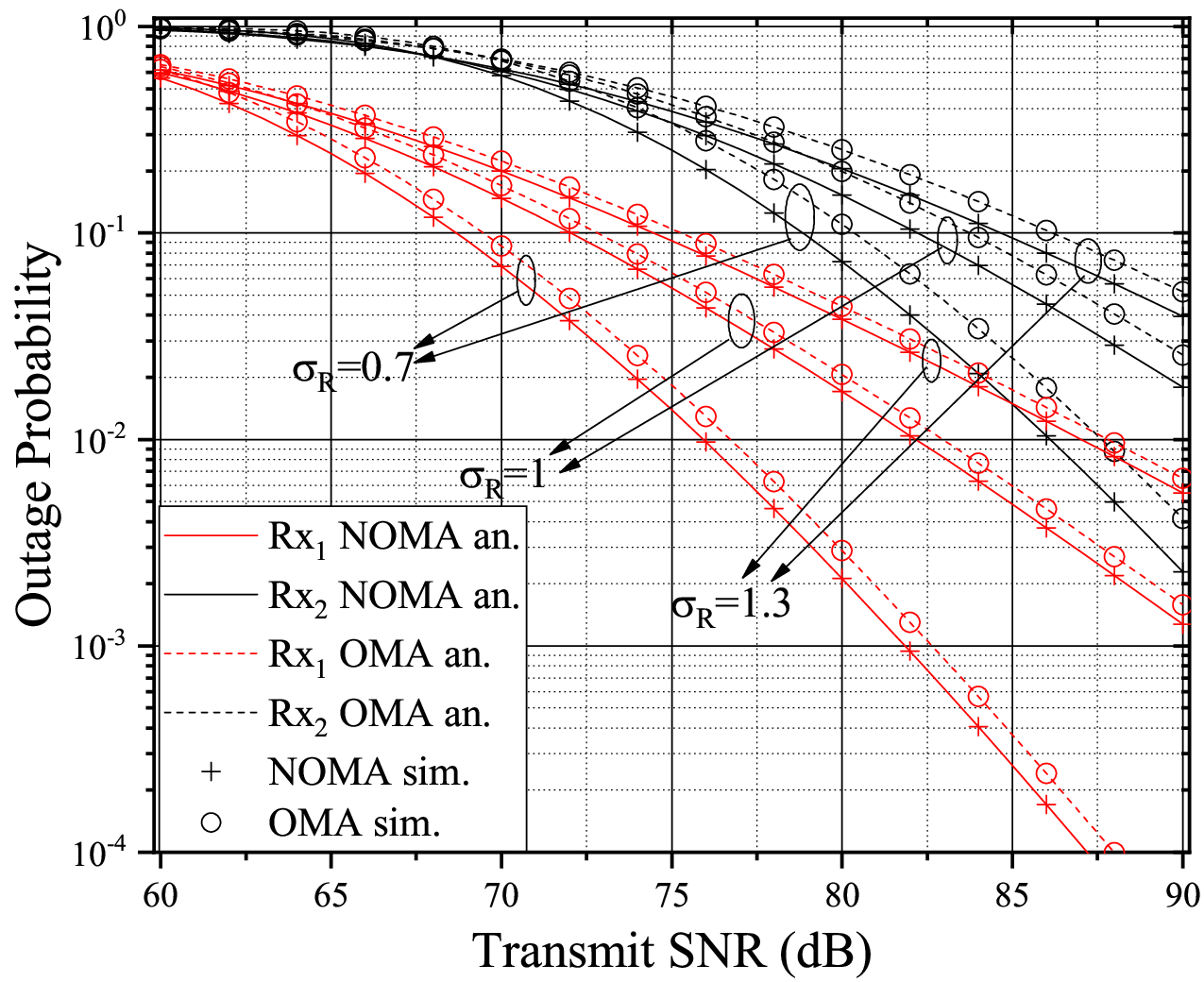}
\caption{OP vs transmit SNR under different $\sigma_R$ values.}
\label{OP_transmitSNR}
\end{figure}

In Fig. \ref{OP_transmitSNR}, Rx1's and Rx2's OP with respect to transmit SNR are depicted for the proposed NOMA scheme under different $\sigma_R$ values. As a benchmark, an OMA scheme is also provided. It is shown that increased transmit SNR leads to decreased OPs for both Rx1 and Rx2. Also, it is obvious that the decrease of $\sigma_R$, which implies the prevalence of more favorable turbulence conditions, results in decreased OPs for both Rx1 and Rx2. Specifically, for a fixed transmit SNR value, for low turbulence fluctuations, i.e., $\sigma_R=0.7$, each of Rx1 and Rx2 achieves a lower OP compared to the case of moderate turbulence conditions, i.e., $\sigma_R=1$, while for strong turbulence fluctuations, i.e., $\sigma_R=1.3$ the worst outage performance for both Rx1 and Rx2 can be noticed. Also, for different $\sigma_R$ values, i.e., for different $\alpha$ and $\beta$ values, the OP curves exhibit different slopes in the high SNR regime, verifying the high SNR insights provided in Remark \ref{rem_3}. Furthermore, it is illustrated that NOMA outperforms OMA for both Rx1 and Rx2 under each of the turbulence conditions, i.e., for all different $\sigma_R$ values, and for the whole range of transmit SNR values. 


\begin{figure}[h]
\centering
\includegraphics[width=0.95\columnwidth]{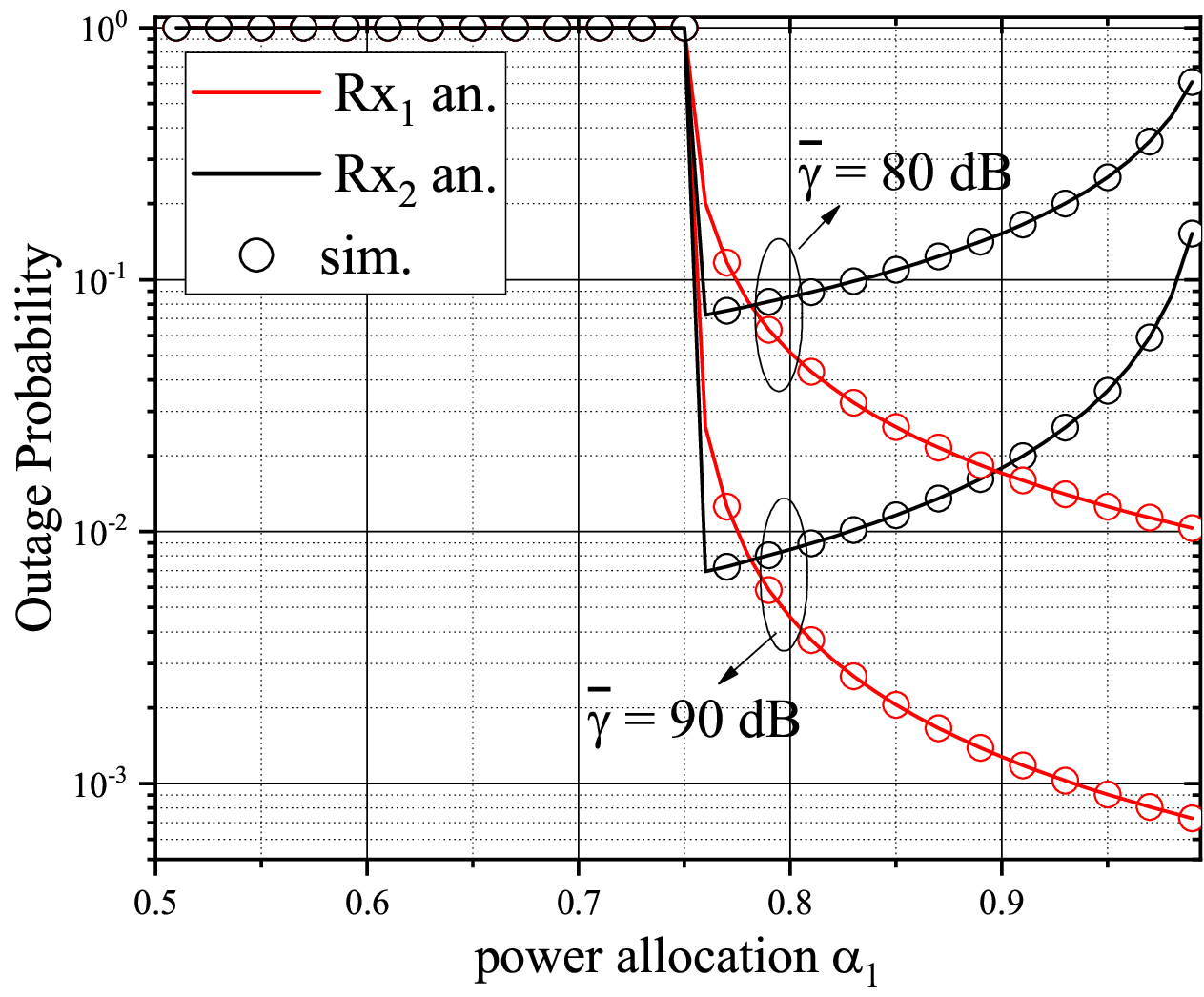}
\caption{OP vs $a_1$ for different $\bar{\gamma}$ values.}
\label{OP_a1}
\end{figure}

\begin{figure}[h]
\centering
\includegraphics[width=0.95\columnwidth]{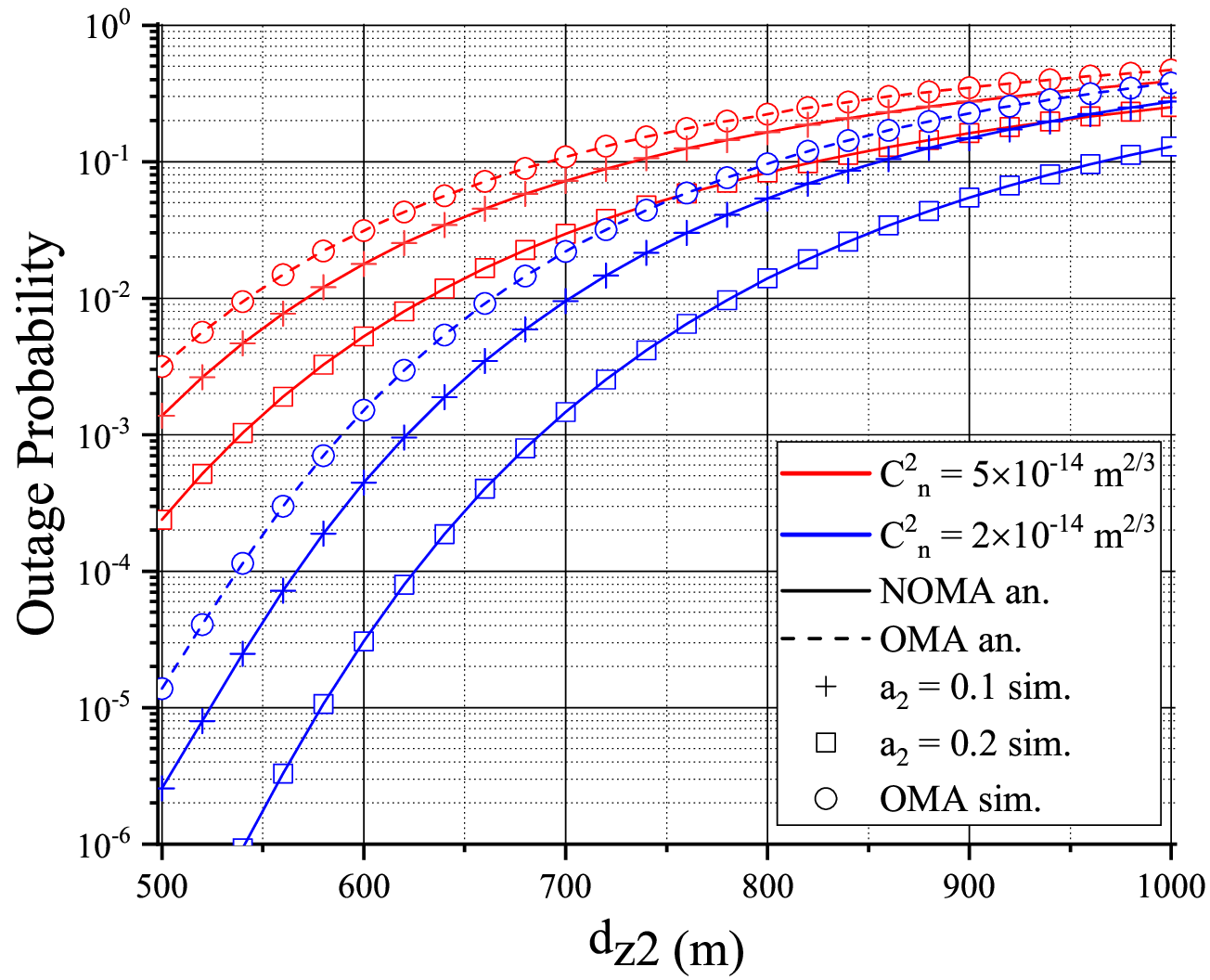}
\caption{OP vs $d_{z2}$ for different $C_n^2$ values.}
\label{OP_dz2}
\end{figure}

Fig. \ref{OP_a1} illustrates the OPs of Rx1 and Rx2 versus the power allocation coefficient $a_1$, evaluated under various values of $\bar{\gamma}$. In the initial range of $a_1$ values, both OPs remain in a constant outage because the second branch in the corresponding equations (\ref{Rx1_OP}), (\ref{Rx2_OP}) remains active, i.e., as also Remarks \ref{rem_Rx1} and \ref{rem_Rx2} highlight, the `operation' condition $\frac{a_1}{a_2}>\gamma^{\text{th}}_{1}$ does not hold. Focusing on Rx1, as $a_1$ increases, its OP decreases. This observation aligns with the impact of $a_1$ on the SINR expression in (\ref{g1}). Specifically, increasing $a_1$ increases the numerator and, simultaneously, reduces $a_2$, i.e., decreases the denominator, leading the resulting ratio in (\ref{g1}) to increase. On the other hand, for Rx2, as $a_1$ increases, its OP initially decreases but subsequently rises, exhibiting a convex trend. To explain this behavior, recall that, as (\ref{U2_out}) clearly shows, successfully decoding message $x_2$ requires for both $\gamma_{21} > \gamma_{1}^{\text{th}}$ and $\gamma_{22} > \gamma_{2}^{\text{th}}$ to hold. From (\ref{g21}), (\ref{g22}) it is evident that increasing $a_1$ enhances $\gamma_{21}$ but reduces $\gamma_{22}$. Accordingly, larger values of $a_1$ increase the probability that $\gamma_{21} > \gamma_{1}^{\text{th}}$ holds, thereby improving the probability of (\ref{U2_out}), as long as $\gamma_{22} > \gamma_{2}^{\text{th}}$ remains relatively likely. However, beyond a certain threshold of $a_1$, further increase of $a_1$ significantly reduce the probability that $\gamma_{22} > \gamma_{2}^{\text{th}}$ holds, negatively impacting the outage performance of Rx2.

In Fig. \ref{OP_dz2}, the OP of Rx2 is provided as a function of its E2E distance for different values of the index of refraction structure, $C_n^2$, with corresponding beam widths $w\left(d_{z2}\right)$ \cite{Najafi_conf}, as well as for different sets of power allocation coefficients $a_1$, $a_2$ when $\bar{\gamma}=80$ dB. We note that, for the extraction of this Fig., we assume that only Rx2's E2E distance ranges and that Rx1's position inside the network remains fixed; thus, Rx1's OP is not illustrated in Fig. 6 since for a fixed E2E distance and under specific power allocation configurations, its OP would remain constant and no further insights would arise. Likewise, increasing the     E2E distance for the farther user, Rx1, while maintaining a fixed distance for Rx2, would yield analogous results due to the similarity of both scenarios. First of all, as expected, for a fixed set of $a_1$, $a_2$ values, an increase in the E2E distance of Rx2 leads to decreased OPs. Additionally, for a fixed E2E distance, as the turbulence conditions deteriorate, i.e., as $C_n^2$ increases, the outage performance of Rx2 becomes worse. Furthermore, under both turbulence conditions represented by different values of $C_n^2$ , NOMA consistently outperforms OMA across the entire E2E transmission distance range. Notably, a discernible trend indicates that as turbulence conditions become less severe, NOMA achieves greater improvements in outage probability, particularly for the Rx2 located nearest to the Tx. Moreover, focusing on the NOMA case, it is observed that the case of $a_2=0.2$ achieves superior outage performance for Rx2 compared to the case of $a_2=0.1$ across all $d_{z2}$ values. This occurs because, as also the SINR derivations in (\ref{g21}) and (\ref{g22}) show, the case of $a_2=0.1$, i.e., $a_1=0.9$, does enhance the ability of Rx2 to successfully decode message $x_1$ and cancel its interference from the received superimposed signal, however, also results in a decrease in the likelihood of successfully decoding $x_2$, even after the removal of $x_1$'s interference. Hence, it becomes obvious that the case of $a_2=0.2$ provides sufficiently high probabilities for both messages to be successfully decoded at 
Rx2, thus, enabling Rx2 to avoid outage. However, at this point, please note that while this power allocation configuration may benefit Rx2, it does not necessarily represent the most fair choice since, as discussed in Fig. \ref{OP_a1}, in terms of Rx1, as large $a_1$ values as possible are preferred in order to ensure the avoidance of an outage event.  


\begin{figure}[h]
\centering
\includegraphics[width= 0.95\columnwidth]{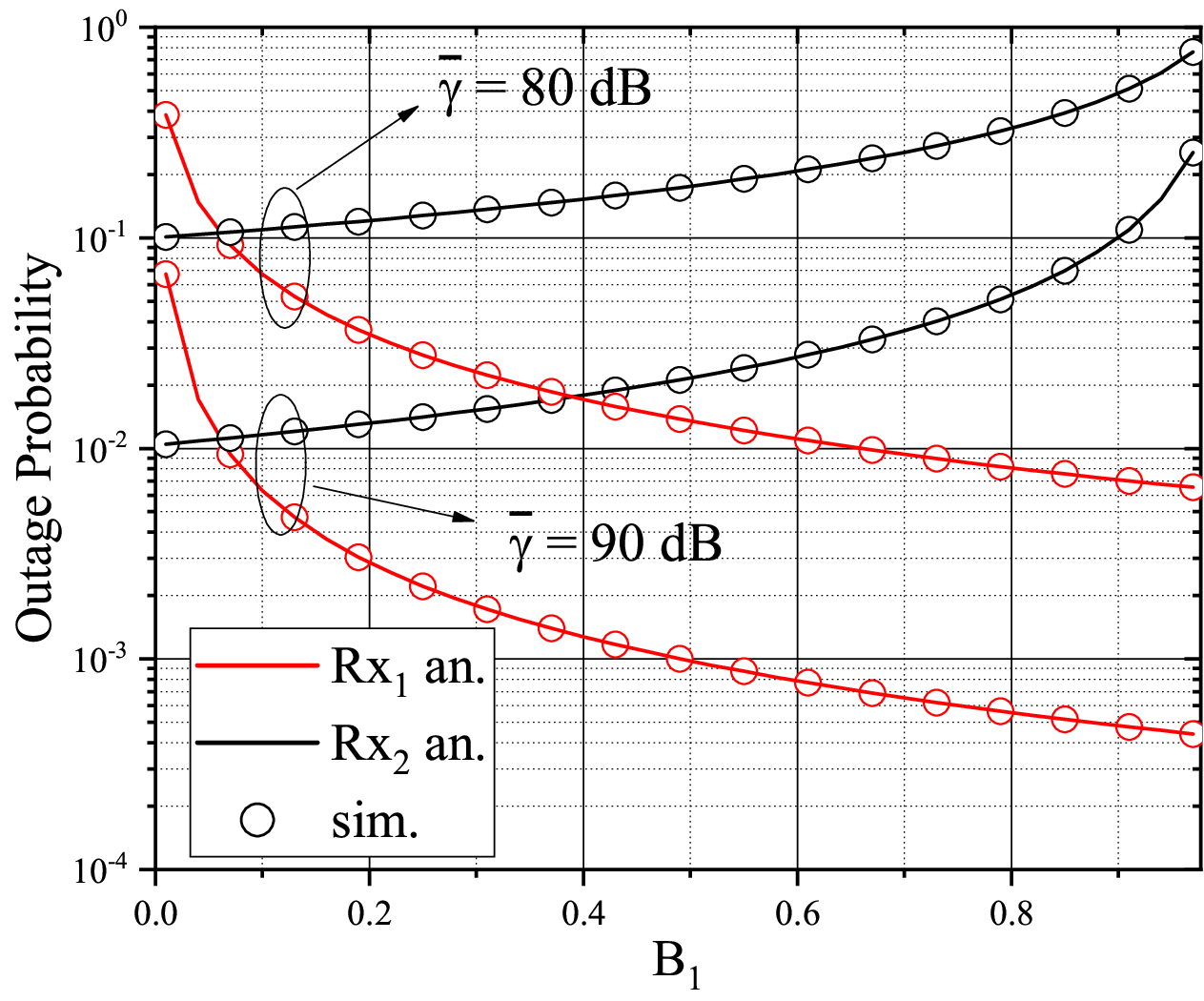}
\caption{OP vs $B_1$ for different $\bar{\gamma}$ values.}
\label{OP_G1}
\end{figure}

In Fig. \ref{OP_G1}, the OPs of Rx1 and Rx2 are plotted with respect to the beam-splitting factor $B_1$ for different $\bar{\gamma}$ values. As expected, higher values of $\bar{\gamma}$ result in improved outage performance for both Rx1 and Rx2 across the entire transmit SNR region. Furthermore, it is evident that increasing $B_1$ allows Rx1 to achieve enhanced outage performance. Recalling that $B_1$ and $B_2$ represent the portions of the incident beam's power at the ORIS assigned to the reflected beams directed toward Rx1 and Rx2, respectively, it is shown that as the reflected beam toward Rx1 receives a greater share of the incident power then its outage performance is improved. This can be explained by observing the impact of a $B_1$ increase on (\ref{g1}). Specifically, it can be seen that the increase of $B_1$ leads to a higher numerator in (\ref{g1}), which is advantageous for the decoding process. However, a $B_1$ increase also contributes to a rise in the denominator of (\ref{g1}), which negatively impacts the SINR of Rx1. Ultimately, given the fact that (\ref{g1}) is an increasing function in terms of $B_1$, the positive impact of the increasing numerator outweighs the negative effect of the increasing denominator, resulting in an overall increase in the SINR for message $x_1$. Consequently, the likelihood of successfully decoding Rx1's message increases, and thus, its OP decreases with higher $B_1$. Conversely, as $B_1$ increases, $B_2$ decreases, which leads to reduced power in the reflected beam directed toward Rx2. It is illustrated that this reduction ultimately results in higher OP values for Rx2. 

\begin{figure}[h]
\centering
\includegraphics[width= 0.95\columnwidth]{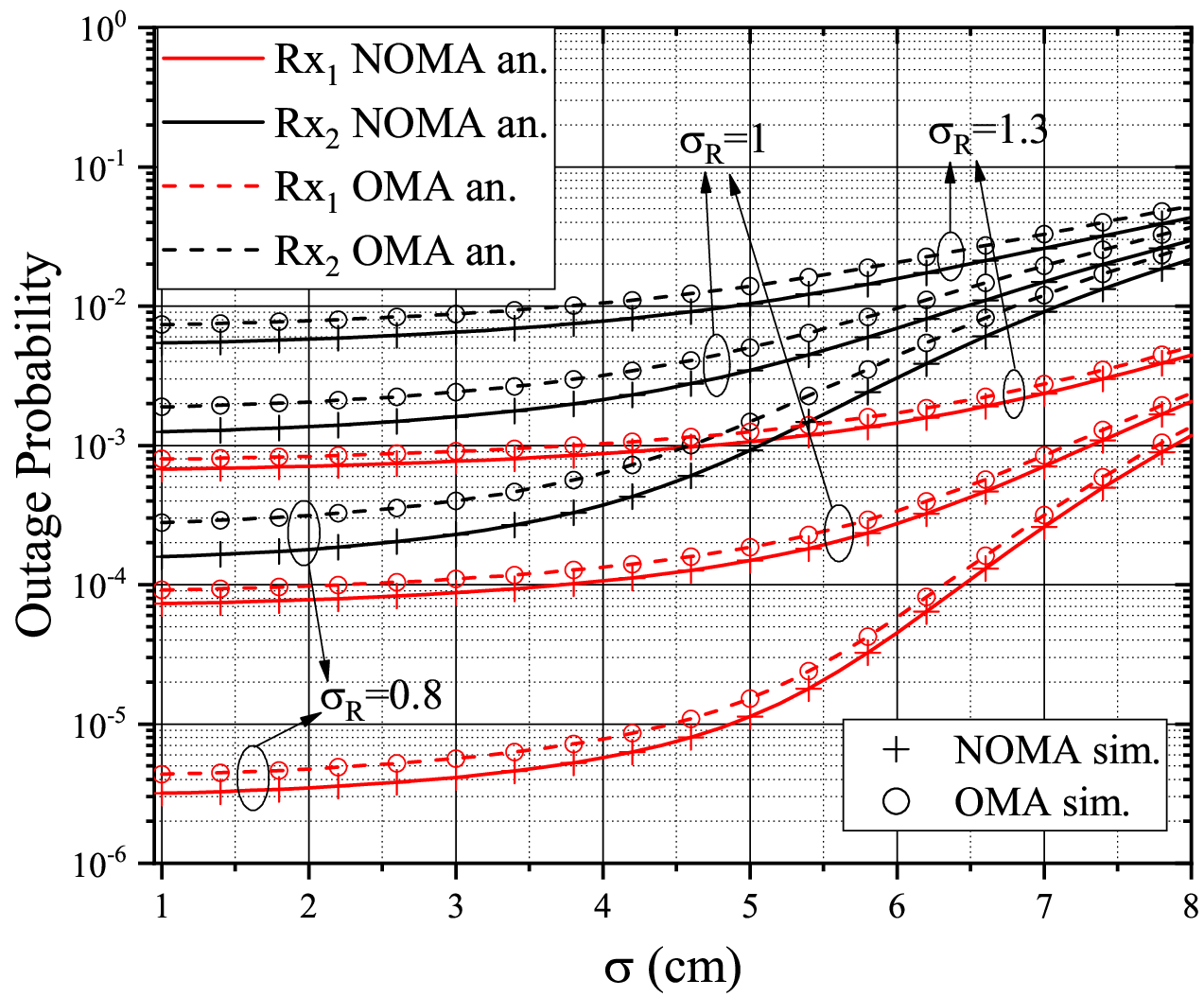}
\caption{OP vs $\sigma_s\!=\!\sigma_r\!=\!\sigma_l\!=\!\sigma$ for different $\sigma_R$ values.}
\label{OP_sigmas}
\end{figure}

In Fig. \ref{OP_sigmas}, the outage performance of both Rx1 and Rx2 is provided as a function of $\sigma_s=\sigma_r=\sigma_l=\sigma$ for different $\sigma_R$ values, when $\bar{\gamma}=100$ dB. Similarly with Fig. \ref{OP_transmitSNR}, it becomes obvious that as $\sigma_R$ increases then both Rx1's and Rx2's OPs increase, highlighting the significant impact of the turbulence conditions on Rx1's, Rx2's outage behavior. Moreover, it is illustrated that NOMA outperforms OMA for both Rx1 and Rx2 across the whole $\sigma$ range of values under all different turbulence conditions. Also, it can be seen that the increase of $\sigma$ leads to an increasing trend for the OPs of both Rx1 and Rx2. However, this increasing trend seems to be heavily depended on the turbulence conditions, i.e., on the $\sigma_R$ values. Specifically, it occurs that for weak turbulence fluctuations, i.e., $\sigma_R=0.8$, increased $\sigma$ values lead to a rapid increase of OP values for both Rx1, Rx2, while for moderate or strong turbulence fluctuations, i.e., $\sigma_R=1$ and $\sigma_R=1.3$, respectively, the increase of $\sigma$ results in a smoother increase of Rx1's and Rx2's OPs. It is also interesting to notice that the increase of the OPs values for both Rx1, Rx2 is not linear across $\sigma$ values. In fact, it occurs that there is a point after which the further increase of $\sigma$ seems to significantly increase the OPs of Rx1, Rx2. This happens when the $\sigma$ begins to achieve higher values compared to the len's diameter $l_d$ of the PDs at the Rxs' side.



\section{Conclusions and future directions}\label{conc}
In this work, we introduced a novel ORIS-based scheme that exploits its beam splitting capabilities to enable NOMA-based multi-user connectivity. To this end, we firstly derived novel analytical expressions, in the form of infinite series, for the E2E channel statistics based on a recently proposed 3D pointing error model. By incorporating these expressions into the OP analysis of a NOMA-based system, we demonstrated its performance under various turbulence conditions and geometric features. Moreover, our high-SNR analysis provided key insights into the asymptotic system behavior, revealing how the physical and design parameters, affect reliability.
Numerical results and simulations validated the accuracy of the derived analytical expressions and offered a comprehensive evaluation of system performance. This evaluation encompassed diverse parameter settings, including transmit SNR, signal power allocation coefficients, beam-splitting factors, fluctuations caused by building sway, environmental conditions, and link distances, demonstrating their impact on system's performance. Also, based on these results, NOMA proved to outperform OMA across a range of system settings. These findings highlight the potential of ORIS-assisted NOMA as a compelling solution for future optical wireless systems, addressing the growing need for enhanced spectral efficiency, massive connectivity, and low-latency communications.

While the reported results provided insights into the integration of ORIS and NOMA in FSO systems, yet several promising directions for future research remain open. Further investigation can extend the proposed framework to an $M$-user scenario, where multi-user connectivity and interference management become more intricate. Another key research avenue is the development of optimal power allocation strategies in the beam-splitting process, aiming to maximize throughput or minimize outage under given resource constraints.

\section*{Appendix A}
At first, utilizing \cite[(8.447/1)]{Gradshteyn2014} and \cite[(8.485)]{Gradshteyn2014} with \cite[(8.445)]{Gradshteyn2014} for $\alpha-\beta \notin \mathbb{Z}$, we can rewrite the Bessel functions $I_0\left(\cdot\right)$ and $K_{\alpha-\beta}\left(\cdot\right)$ of (\ref{f_hg}) and (\ref{f_hs}), respectively, in the following forms 
\begin{equation}
    I_0\left(v \ln \left(\frac{h_s}{\chi}\right)\right)=\sum_{k=0}^{\infty} \frac{\left(\frac{v}{2}\right)^{2k} \ln^{2k}(\frac{h_s}{\chi})}{(k!)^2},
    \label{I_0}
\end{equation}
\begin{equation}
    K_{\alpha-\beta}\left(2 \sqrt{\alpha \beta h_s}\right)=\frac{1}{2} \mathrm{G}_{0,2}^{2,0}\left(\alpha \beta h_s \left\lvert \begin{array}{c}- \\ \frac{\alpha-\beta}{2} ,\frac{\beta-\alpha}{2}\!\!\!\end{array}\right.\right),
    \label{K_0}
\end{equation}
where we denote $\chi = \frac{h}{A_0 h_l}$ and $v = \frac{\left(1-q^2\right) \omega}{2q}$ for convenience. By utilizing (\ref{I_0}) and (\ref{K_0}), as well as interchanging the order of summation and integration, which is justified by the existence of the PDF \footnote{Both $f_{h_s}(h_s)$ and $f_{h_g}(h_g)$ are well-defined and integrable over their respective domains \cite{Najafi_new}} in (\ref{f_h}), we get
\begin{align}
    f_h\left(h\right) &= \mathcal{D} \chi^{c-1} \sum_{k=0}^{\infty} \frac{\left(\frac{v}{2}\right)^{2k}}{(k!)^2} \int_{\chi}^{\infty}  h_s^{\frac{a+b}{2}-c-1}\ln^{2k}\left(\frac{h_s}{\chi}\right) 
    \nonumber\\ & \times \mathrm{G}_{0,2}^{2,0}\left(\alpha \beta h_s \mathrel{\Big|} \begin{array}{c}- \\ \!\!\! \frac{\alpha-\beta}{2} ,\frac{\beta-\alpha}{2}\!\!\!\end{array}\right) dh_s,
    \label{f_h_v2}
\end{align}
where 
\begin{equation}
    \mathcal{D} = \frac{\omega\left(\alpha \beta\right)^{\frac{\alpha+\beta}{2}} }{A_0 h_l \Gamma(\alpha)\Gamma(\beta)}.
\end{equation}

\begin{figure}[t]
\centering
	\includegraphics[scale = 1.05]{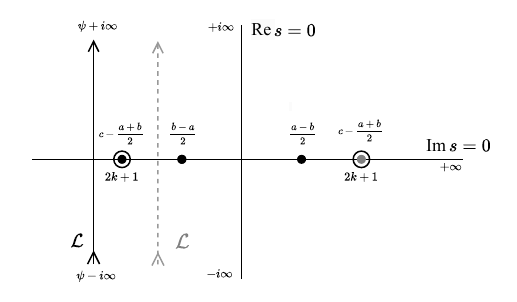}
 \vspace{-0.25cm}	
 \caption{Schematic representation of the Meijer-G functions integration line $\mathcal{L}$ of \eqref{eq:cdf} in the complex domain, in relation with the value cases of the system parameters.}
	\label{Fig:path}
 \vspace{-0.5cm}
\end{figure}

Next, using the definition of Meijer-G function through Mellin–Barnes integral \cite[(9.301)]{Gradshteyn2014}, we can express the Meijer-G function of (\ref{f_h_v2}) as
\begin{align}
    &\mathrm{G}_{0,2}^{2,0} \left(\alpha \beta h_s \mathrel{\Big|} \begin{array}{c}- \\ \!\!\! \frac{\alpha-\beta}{2} ,\frac{\beta-\alpha}{2}\!\!\!\end{array}\right) = \frac{1}{2\pi i} \int_{\mathcal{L}} \Gamma\left(\frac{\alpha-\beta}{2}-s\right)
    \nonumber\\ & \times \Gamma\left(\frac{\beta-\alpha}{2}-s\right) \left(\alpha \beta h_s\right)^s ds.
    \label{MeijerG_repr}
\end{align}
By firstly substituting (\ref{MeijerG_repr}) in (\ref{f_h_v2}) and then interchanging the order of integration between the two integrals, it yields
\begin{align}
    f_h\left(h\right) &=\mathcal{D} \chi^{c-1} \sum_{k=0}^{\infty} \frac{\left(\frac{v}{2}\right)^{2k}}{(k!)^2}  
    \nonumber\\ & \times \frac{1}{2 \pi i} \int_\mathcal{L} \Gamma\left(\frac{\alpha-\beta}{2}-s\right) \Gamma\left(\frac{\beta-\alpha}{2}-s\right) (\alpha \beta )^s  \nonumber\\ & \times \underbrace{\left[\int_\chi^{\infty} h_s^{s+\frac{\alpha+\beta}{2}-c-1} \ln^{2 k}\left(\frac{h_s}{\chi}\right) d h_s\right]}_{\text{A}} ds.
    \label{f_h_v3}
\end{align}
Consequently, if we perform the change of variables $y = \ln\left(h_s/\chi\right) $ the underlined integral of (\ref{f_h_v3}) can be evaluated as: 
\begin{align}
    A &= \chi^{s + \frac{\alpha+\beta}{2}-c }\int_0^{\infty} e^{y\left(s+\frac{\alpha+\beta}{2}-c\right)} y^{2 k} d y 
    \nonumber\\ &  = \frac{\Gamma(2 k+1)\, \chi^{s + \frac{\alpha+\beta}{2}-c}}{\left(c-s-\frac{\alpha+\beta}{2}\right)^{2 k+1}}.
    \label{A}
\end{align}
Note that the integral of \eqref{A} always converges, regardless of the values of the parameters $a$, $b$ and $c$. This is attributed to the integration path $\mathcal{L}$, which 
is chosen to lie along the left side of all the poles. Specifically the contour $\mathcal{L}$  is a straight line parallel to the imaginary axis which runs from $s = \psi -i\infty$  to $s = \psi + i\infty$ where $\psi< c - \frac{a+b}{2}$ is a real constant \cite{Luke, Fox-H} ensuring the convergence of \eqref{A}. The integration line paths $\mathcal{L}$ along with the poles of \eqref{poles} can be seen at Fig. \ref{Fig:path}.
Additionally, from the well-known  property of Gamma function, 
\begin{equation}
    \Gamma(x+1)=x \Gamma(x), \:\:\:x \in \mathbb{Z},
    \label{prop}
\end{equation}
it holds 
\begin{align}
    \left(c-s-\frac{\alpha+\beta}{2}\right)^{2 k+1} = \frac{\Gamma^{2k+1}\left(c-s-\frac{\alpha+\beta}{2}\right)}{\Gamma^{2k+1}\left(c-s-\frac{\alpha+\beta}{2}+ 1\right)},
\end{align}
and thus, (\ref{f_h_v3}) can be transformed into 
\begin{align}
    &f_h\left(h\right) =\mathcal{D} \chi^{\frac{\alpha+\beta}{2}-1} \sum_{k=0}^{\infty} \frac{(2k)!\left(\frac{v}{2}\right)^{2k}}{(k!)^2} \frac{1}{2 \pi i}\int_{\mathcal{L}} (\alpha \beta \chi)^s
    \nonumber\\ & \times\!\frac{\Gamma\left(\frac{\alpha-\beta}{2}-s\right) \Gamma\left(\frac{\beta-\alpha}{2}-s\right)\Gamma^{2k+1}\!\left(c-s-\frac{\alpha+\beta}{2}\right)}{\Gamma^{2k+1}\!\left(c-s-\frac{\alpha+\beta}{2}+ 1\right)} ds.
    \label{poles}
\end{align}
The integral in (\ref{poles}) resembles the definition of Meijer-G function through the Mellin-Barnes integral and as a result (\ref{poles}) becomes
\begin{align}
    f_h\left(h\right) &=\mathcal{D} \chi^{\frac{\alpha+\beta}{2}-1} \sum_{k=0}^{\infty} \frac{(2k)!\left(\frac{v}{2}\right)^{2k}}{(k!)^2} 
    \nonumber\\ & \times\mathrm{ G}_{2k+1,2k+3}^{2k+3,0}\!\left(\!\alpha \beta \chi \! \mathrel{\Bigg|} \begin{array}{c}
	    \left\{c - \frac{\alpha+\beta}{2}+1\right\}_{1}^{2k+1} \\
	    \!\!\!\! \frac{\alpha-\beta}{2}, \frac{\beta-\alpha}{2}, \left\{c-\frac{\alpha+\beta}{2}\right\}_{1}^{2k+1}\!\!\!\!
	\end{array}\!\right)\,.
    \label{f_h_v4}
\end{align}
Finally, by substituting $\mathcal{D}$ and $\chi$ in (\ref{f_h_v4}) and then utilizing \cite[(9.31.5)]{Gradshteyn2014}, \eqref{eq:pdf} occurs.

\section*{Appendix B} 

Due to the structure of \eqref{eq:cdf}, it is not possible to directly apply \cite[(9.304)]{Gradshteyn2014}, as certain parameters within the Meijer-G function are identical. In our case, we have a pole at $s=c$ of order $2k+1$ that is represented by the parameters $b_3,\dots,b_{2k+3}$. In this scenario, we utilize the result provided in, \cite[Theorem 1.12]{H-transforms} or \cite[Theorem 2]{Fox-H} in order to express the Meijer-G function\footnote{To apply the expressions, we first convert the Meijer-G function into a Fox-H function by utilizing the straightforward identity in \cite[(6.2.8)]{the_alg}.} of \eqref{out_p} as:

\begin{align}
    &\mathrm{G}_{2k+2,2k+4}^{2k+3,1}\left( z \mathrel{\bigg|} \begin{array}{c}
		\!\!a_1,\dots, a_{2k+2} \\
	    \!\! b_1,\dots, b_{2k+4} \!\!
	\end{array}\right) \stackrel{z \rightarrow 0}{\cong} \nonumber \\ & \sum_{\substack{j=1\\j\neq 3,\dots,2k+3}}^{2k+3} g^{*}_{j} z^{b_j} + \mathcal{G}^{*}_{k,c} \left(\ln(z)\right)^{2k} z^c
\end{align}
where 
\begin{equation}
    g^{*}_{j} = \frac{\displaystyle \prod_{i=1}^{2k+3} \Gamma\left(b_i-b_j\right) \displaystyle \Gamma\left(b_j\right)}{\displaystyle \prod_{i=2}^{2k+2} \Gamma\left(a_i-b_j \right) \displaystyle  \Gamma\left(1+b_j\right)} 
    \label{Meijerg}
\end{equation}
and 
\begin{align}
&\mathcal{G}^{*}_{k,c} = \frac{\left(-1\right)^{2k}}{\left(2k\right)!} \left\{ \prod_{r=1}^{2 k+1} \frac{(-1)^{l_r}}{l_{r}!}\right\} \nonumber \\ & \times \frac{ \displaystyle \prod_{\substack{i=1 \\ i \neq 3, \cdots, 2k+3}}^{2k+3} \Gamma\left(b_i-c\right) \Gamma\left(c\right)}{\displaystyle \prod_{i=2}^{2k+4} \Gamma\left(a_i-c\right)   \Gamma\left(1+c\right)},
\label{G_K}
\end{align}
where the first sum refers to the simple poles of the Meijer-G function and the second term to the pole $s=c$ of order $2k+1$. Note that the asymptotic behavior of \eqref{Meijerg} for small values of $z \rightarrow 0$ is determined solely by the parameters $b_1,\dots,b_{m=2k+3}$, due to the condition $\Delta = q-p = 2>0$ and also according to the \cite[Corollary 1.12.1]{H-transforms} the principal term in the asymptotic expansion is the one with the minimum value parameter, in our case $\min\{\alpha,\beta,c\}$.
The indices $l_r\in \{0,1,2,\dots\}$ in \eqref{G_K} arise from the repeated-derivative expansions that are used to compute the residue of the high-order pole \cite[(1.4.2)]{H-transforms}. Because in the integral representation of \eqref{eq:cdf}, we have $(2 k+1)$ identical factors $\Gamma(c-s)$, the $\left\{l_1, \ldots, l_{2 k+1}\right\}$ must satisfy $l_1+l_2+\cdots+l_{2 k+1}=2 k$,
so that the total order of differentiation is $(2 k)$. Considering the following well-known generating function
\begin{equation}
    A(x)=\sum_{l=0}^{\infty} \frac{(-x)^l}{l!}=e^{-x},
\end{equation}
we express the product as
\begin{align}  
e^{-\left(2 k+1\right)x}&=\left(\sum_{l=0}^{\infty} \frac{(-x)^l}{l!}\right)^{2 k+1} = \sum_{t=0}^{\infty} \frac{\left(-\left(2 k+1\right) x\right)^t}{t!} \nonumber \\
&=\sum_{t=0}^{\infty}\left(\sum_{l_1+l_2+\cdots+l_{2 k+1}=t} \prod_{r=1}^{2 k+1} \frac{(-x)^{l_r}}{l_{r}!}\right).
\end{align}

For the case of $t=2k$, equating the coefficients on both sides and setting $x=1$, we get
\begin{equation}
    \sum_{l_1+l_2+\cdots+l_{2 k+1}=2 k} \prod_{r=1}^{2 k+1} \frac{(-1)^{l_r}}{l_{r}!}=\frac{(2 k+1)^{2 k}}{(2 k)!}
    \label{sum_j}
\end{equation}

Substituting \eqref{sum_j} in \eqref{G_K} and calculating the products of Gamma functions in \eqref{Meijerg} and \eqref{G_K}, we derive \eqref{Meijer_asy}, given at the top of the next page. In the final step, we substitute \eqref{Meijer_asy} into \eqref{eq:cdf}, while also accounting for the case $k=0$, in which the pole at $s=c$ is simple. This concludes the proof.

\begin{figure*} 
\begin{align}
    &\mathrm{ G}_{2k+2,2k+4}^{2k+3,1}\left(z \mathrel{\bigg|} \begin{array}{c}
		\!\!1, \left\{c+1\right\}_{1}^{2k+1} \\
	    \!\! \alpha, \beta, \left\{c\right\}_{1}^{2k+1}\!, 0 \!\!
	\end{array}\right) \stackrel{z \rightarrow 0}{\sim} \frac{\Gamma\left(\beta-\alpha\right)}{\alpha \left(c-\alpha\right)^{2k+1}} z^\alpha + \frac{\Gamma\left(\alpha-\beta\right)}{\beta \left(c-\beta\right)^{2k+1}} z^\beta \nonumber \\ 
    & +\frac{\left(2k+1\right)^{2k}\Gamma\left(\alpha-c\right)\Gamma\left(\beta-c\right)}{c \left(2k\right)!^2} \left(\ln z\right)^{2k} z^c. 
    \label{Meijer_asy}
\end{align}
\hrulefill
\end{figure*}

\section*{Appendix C}
Considering that the asymptotic expression of the CDF for values close to zero, does not necessarily adhere to the distribution properties of the corresponding CDF given in (\ref{eq:cdf}), we need to re-examine the convergence of the asymptotic expression in (\ref{asympt}). Specifically, substituting \eqref{Meijer_asy} in \eqref{eq:cdf}, the following sum arises in the case of $\min\left(\alpha,\beta\right)<c$:
\begin{align}
    S = \sum_{k=0}^{\infty} \binom{2k}{k}\left(\frac{v}{2\left(c-\min\left(\alpha,\beta\right)\right)}\right)^{2k}.
    \label{conv}
\end{align}
The sum in \eqref{conv} converges iff $v<c-\min\left(\alpha,\beta\right)$ \cite[(0.241/3)]{Gradshteyn2014}. Of note, this condition holds true for a majority of physical scenarios and geometric characteristics of the ORIS size and placement \cite{Najafi_new}, especially for medium and high turbulence conditions. In that case, utilizing \cite[(0.241/3)]{Gradshteyn2014}, \eqref{conv} can be expressed as: 
\begin{align}
    S = \frac{c-\min\left(\alpha,\beta\right)}{\sqrt{\left(c-\min\left(\alpha,\beta\right)\right)^2-v^2}}
\end{align}

On the other hand, when $c<\min\left(\alpha,\beta\right)$, the asymptotic expression is given by the second branch of \eqref{asympt}. For the second branch, convergence of the infinite sum can be straightforwardly demonstrated for every value of $v,c,\alpha,\text{ and }\beta$ by utilizing the Stirling approximation for the factorial terms and applying the Ratio test. However, in the high-SNR regime, the logarithm can attain large values, which, for certain parameter sets, may require a large number of terms for the summation to converge accurately to multiple decimal places. In such cases, a heuristic approximation can be applied by considering only the first few terms, yielding meaningful results that approximate the exact solution.


\balance

\bibliographystyle{IEEEtran}
\bibliography{IEEEabrv,References_2}
\end{document}